\documentclass[12pt,a4paper]{article}
\usepackage[dvips]{graphicx}

\makeatletter
\newif\if@preliminary
\@preliminaryfalse
\def\preliminary{\@preliminarytrue}
\def\preprintno#1{\def\@preprintno{#1}}
\def\address#1{\def\@address{#1}}

\def\abstract#1{\def\@abstract{#1}}
\renewcommand\abstractname{ABSTRACT}
\newlength\preprintnoskip
\newlength\abstractwidth
\@titlepagetrue
\renewcommand\maketitle{\begin{titlepage}%
  \let\footnotesize\small
  \hfill\parbox{\preprintnoskip}{%
  \begin{flushright}\@preprintno\end{flushright}}\hspace*{1cm}
  \vskip 60\p@
  \begin{center}%
    {\Large\bf\boldmath \@title \par}\vskip 1cm%
    {\sc\@author \par}\vskip 3mm%
    {\@address \par}%
    \if@preliminary
      \vskip 2cm {\large\sf PRELIMINARY DRAFT \par \@date}%
    \fi
  \end{center}\par
  \@thanks
  \vfill
  \begin{center}%
    \parbox{\abstractwidth}{\centerline{\abstractname}%
    \vskip 3mm%
    \@abstract}
  \end{center}
  \end{titlepage}%
  \setcounter{footnote}{0}%
  \let\thanks\relax\let\maketitle\relax
  \gdef\@thanks{}\gdef\@author{}\gdef\@address{}%
  \gdef\@title{}\gdef\@abstract{}\gdef\@preprintno{}
}%
\def\shortletter{%
  \setcounter{secnumdepth}{5}
  \def\paragraph{%
    \@startsection{paragraph}{4}{\parindent}%
      {3.25ex \@plus1ex \@minus.2ex}{-.5em}%
      {\reset@font\normalsize\bfseries}}%
  \renewcommand\theparagraph{\arabic{paragraph}.\hskip-.5em}
  \def\subparagraph{%
    \@startsection{subparagraph}{5}{\parindent}%
      {3.25ex \@plus1ex \@minus.2ex}{-.5em}%
      {\reset@font\normalsize\bfseries}}%
  \renewcommand\thesubparagraph{(\alph{subparagraph})\hskip-.5em}
}
\def\thesection{\arabic{section}.}

\def\appendix{\setcounter{section}{0}
 \def\thesection{Appendix \Alph{section}:}
 \def\theequation{\Alph{section}.\arabic{equation}}}
\def\@citex[#1]#2{\if@filesw\immediate\write\@auxout{\string\citation{#2}}\fi
  \def\@citea{}\@cite{\@for\@citeb:=#2\do
    {\@citea\def\@citea{,\penalty\@m}\@ifundefined
       {b@\@citeb}{{\bf ?}\@warning
       {Citation `\@citeb' on page \thepage \space undefined}}%
\hbox{\csname b@\@citeb\endcsname}}}{#1}}
\def\citerange{\@ifnextchar [{\@tempswatrue\@citexr}{\@tempswafalse\@citexr[]}}
\def\@citexr[#1]#2{\if@filesw\immediate\write\@auxout{\string\citation{#2}}\fi
  \def\@citea{}\@cite{\@for\@citeb:=#2\do
    {\@citea\def\@citea{--\penalty\@m}\@ifundefined
       {b@\@citeb}{{\bf ?}\@warning
       {Citation `\@citeb' on page \thepage \space undefined}}%
\hbox{\csname b@\@citeb\endcsname}}}{#1}}
\long\def\@makecaption#1#2{%
  \vskip\abovecaptionskip
  \sbox\@tempboxa{#1: \emph{#2}}%
  \ifdim \wd\@tempboxa >\hsize
    #1: \emph{#2}\par
  \else
    \hbox to\hsize{\hfil\box\@tempboxa\hfil}%
  \fi
  \vskip\belowcaptionskip}
\def\fmslash{\@ifnextchar[{\fmsl@sh}{\fmsl@sh[0mu]}}
\def\fmsl@sh[#1]#2{%
  \mathchoice
    {\@fmsl@sh\displaystyle{#1}{#2}}%
    {\@fmsl@sh\textstyle{#1}{#2}}%
    {\@fmsl@sh\scriptstyle{#1}{#2}}%
    {\@fmsl@sh\scriptscriptstyle{#1}{#2}}}
\def\@fmsl@sh#1#2#3{\m@th\ooalign{$\hfil#1\mkern#2/\hfil$\crcr$#1#3$}}
\makeatother

\def\fmfL(#1,#2,#3)#4{\put(#1,#2){\makebox(0,0)[#3]{#4}}}

\def\hc{\mbox{h.c.}}
\def\op{{\cal O}}

\begin{document}
\shortletter        
\baselineskip20pt   
\preprintno{DESY 95--217}
\title{%
 ANOMALOUS COUPLINGS\\
 IN THE HIGGS-STRAHLUNG PROCESS
}
\author{%
 W.~Kilian,
 M.~Kr\"amer,
 and P.M.~Zerwas
}
\address{%
 Deutsches Elektronen-Synchrotron DESY\\
 D-22603 Hamburg/FRG
}
\abstract{%
  The angular distributions in the Higgs-strahlung process $e^+e^-\to
  HZ\to H\bar f f$ are uniquely determined in the Standard Model.  We
  study how these predictions are modified if non-standard couplings
  are present in the $ZZH$ vertex, as well as lepton-boson contact
  terms.  We restrict ourselves to the set of operators which are
  singlets under standard $SU_3\times SU_2\times U_1$ transformations,
  CP conserving, dimension 6, helicity conserving, and custodial
  $SU_2$ conserving. 
}
\maketitle

\paragraph{}
The Higgs-strahlung process~\cite{hst}
\begin{equation}\label{hst}
  e^+e^-\to HZ \to H\bar f f
\end{equation}
together with the $WW$ fusion process, are the most important
mechanisms for the production of Higgs bosons in $e^+e^-$
collisions~\cite{LEP2,FLC}.  Since the $ZZH$ vertex is uniquely
determined in the Standard Model (SM), the production cross section of
the Higgs-strahlung process, the angular distribution of the $HZ$
final state as well as the fermion distribution in the $Z$ decays can
be predicted if the mass of the Higgs boson is fixed~\cite{BZ}.
Deviations from the pointlike coupling can occur in models with
non-pointlike character of the Higgs boson itself or through interactions
beyond the SM at high energy scales.  We need not specify the
underlying theory but instead we will adopt the usual assumption that
these effects can globally be parameterized by introducing a set of
dimension-6 operators
\begin{equation}\label{L-eff}
  {\cal L} = {\cal L}_{\rm SM} + \sum_i\frac{\alpha_i}{\Lambda^2}\op_i
\end{equation}
The coefficients are in general expected to be of the order
$1/\Lambda^2$, where $\Lambda$ denotes the energy scale of the new
interactions.  However, if the underlying theory is weakly
interacting, the $\alpha_i$ can be significantly smaller than unity,
in particular for loop-induced operators.  [It is assumed \emph{a
priori} that the ratio of the available c.m.\ energy to $\Lambda$ is
small enough for the expansion in powers of $1/\Lambda$ to be
meaningful.]

If we restrict ourselves to operators~\cite{BW85} which are singlets
under $SU_3\times SU_2\times U_1$ transformations of the SM gauge
group, CP conserving, and conserving the custodial $SU_2$ symmetry, the
following bosonic operators are relevant for the Higgs-strahlung
process:
\begin{eqnarray}
  \op_{\partial\varphi} 
   &=& \frac12|\partial_\mu(\varphi^\dagger\varphi)|^2\\
  \op_{\varphi W}
   &=& \frac12\varphi^\dagger \vec W_{\mu\nu}^2\varphi \\
  \op_{\varphi B}
   &=& \frac12\varphi^\dagger B_{\mu\nu}^2\varphi 
\end{eqnarray}
where the gauge fields $W^3,B$ are given by the $Z,\gamma$ fields.
This set of operators is particularly interesting because it does not
affect, at tree level, observables which do not involve the Higgs
particle explicitly.  [It is understood that the fields and parameters
are (re-)normalized in the Lagrangian ${\cal L}$ in such a way that
the particle masses and the electromagnetic coupling retain their
physical values.]

In addition, we consider the following helicity-conserving fermionic
operators which induce contact terms contributing to $e^+e^-\to ZH$:
\begin{eqnarray}
  \op_{L1}
   &=& (\varphi^\dagger iD_\mu\varphi)
       (\bar\ell_L\gamma^\mu\ell_L) + \hc \\
  \op_{L3}
   &=& (\varphi^\dagger \tau^a iD_\mu\varphi)
       (\bar\ell_L\tau^a\gamma^\mu\ell_L) + \hc \\
  \op_R
   &=& (\varphi^\dagger iD_\mu\varphi)
       (\bar e_R\gamma^\mu e_R) + \hc 
\end{eqnarray}
[$\ell_L$ and $e_R$ denote the left-handed lepton doublet and the
right-handed singlet, respectively.  The vacuum expectation value of
the Higgs field is given by $\langle\varphi\rangle = (0,v/\sqrt2)$
with $v=246\,{\rm GeV}$, and the covariant derivative acts on the
Higgs doublet as $D_\mu = \partial_\mu - \frac{i}{2}g\tau^a W^a_\mu +
\frac{i}{2}g'B_\mu$.]  Helicity-violating fermionic operators do not
interfere with the SM amplitude, so that their contribution to the
cross section is suppressed by another power of~$\Lambda^2$.  The
helicity-conserving fermionic operators modify the SM $Zee$ couplings
and are therefore constrained by the measurements at LEP1; however, it
is possible to improve on the existing limits by measuring the
Higgs-strahlung process at a high-energy $e^+e^-$ collider since the
impact on this process increases with energy~\cite{GW}.

\begin{figure}[bth]
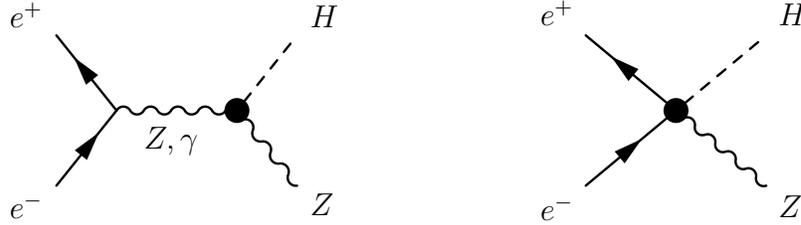

\unitlength 1mm
\begin{center}
\begin{picture}(40,20)
  \put(0,0){\includegraphics{fmgraphs.1}}
  \input{fmgraphs.t1}
\end{picture}\hspace{3cm}
\begin{picture}(30,20)
  \put(0,0){\includegraphics{fmgraphs.2}}
  \input{fmgraphs.t2}
\end{picture}
\end{center}
\caption{Anomalous $ZZH$/$\gamma ZH$ couplings and $e^+e^-ZH$
contact terms in the Higgs-strahlung process.}
\label{ZZH}
\end{figure}

The effective $ZZH$ and the induced $\gamma ZH$ interactions
(Fig.\ref{ZZH}, left diagram) may be written
\begin{eqnarray}\label{L-ZZH}
  {\cal L}_{ZZH} &=& g_ZM_Z\left(
        \frac{1+a_0}{2} Z_\mu Z^\mu H 
        + \frac{a_1}{4} Z_{\mu\nu}Z^{\mu\nu}H\right)\\
  {\cal L}_{\gamma ZH} &=&
        g_ZM_Z\, \frac{b_1}{2}  Z_{\mu\nu}A^{\mu\nu}H
\end{eqnarray}
where $g_Z=M_Z\,\sqrt{4\sqrt{2}G_F}$.  Additional operators $Z_\mu
Z^{\mu\nu}\partial_\nu H$ and $Z_\mu A^{\mu\nu}\partial_\nu H$ are
redundant in this basis: They may be eliminated in favor of the other
operators and the contact terms by applying the equations of motion.
The remaining coefficients are given by
\begin{eqnarray}
  \label{a0'}
  a_0 &=& -\frac{1}{2}\alpha_{\partial\varphi}\, v^2 / \Lambda^2\\
  a_1 &=& 4g_Z^{-2}
        \left( c_W^2\alpha_{\varphi W} + s_W^2\alpha_{\varphi B}\right) /
        \Lambda^2\\
  b_1 &=& 4g_Z^{-2}
        c_Ws_W\left( -\alpha_{\varphi W} + \alpha_{\varphi B}\right) /
        \Lambda^2
\end{eqnarray}
where $s_W$ and $c_W$ are the sine and cosine of the weak mixing
angle, respectively.

In the same way the $e\bar eHZ$ contact interactions (Fig.\ref{ZZH},
right diagram) can be defined for left/right-handed electrons and
right/left-handed positrons
\begin{equation}\label{L-contact}
  {\cal L}_{eeZH} 
  = g_ZM_Z\left[
   c_L \bar e_L\fmslash Z e_L H + c_R \bar e_R\fmslash Z e_R H
   \right]
\end{equation}
with
\begin{eqnarray}
  c_L &=& -2g_Z^{-1}  \left(\alpha_{L1}  + \alpha_{L3}\right)/
  \Lambda^2\\   
  c_R &=& -2g_Z^{-1}  \alpha_R / \Lambda^2
\end{eqnarray}

Some consequences of these operators for Higgs production in $e^+e^-$
collisions have been investigated in the past.  Most recently, the
effect of novel $ZZH$ vertex operators and $\ell\bar\ell ZH$ contact
terms on the total cross sections for Higgs production has been
studied in Ref.\cite{GW}.  The impact of vertex operators on angular
distributions has been analyzed in Refs.\cite{SH} and~\cite{GR}.  We
expand on these analyses by studying the angular distributions for the
more general case where both novel vertex operators and contact
interactions are present.  The analysis of angular distributions in
the Higgs-strahlung process~(\ref{hst}) allows us to discriminate
between various novel interactions.  In fact, the entire set of
parameters $a_0,a_1,b_1$ and $c_L,c_R$ can be determined by measuring
the polar and azimuthal angular distributions as a function of the
beam energy if the electron/positron beams are unpolarized. As
expected, the energy dependence of the polar angular distribution is
sufficient to provide a complete set of measurements if longitudinally
polarized electron beams are available\footnote{Since we can restrict
  ourselves to helicity-conserving couplings, as argued before,
  additional positron polarization need not be required.}.

\begin{figure}[bt]
\begin{center}
\includegraphics{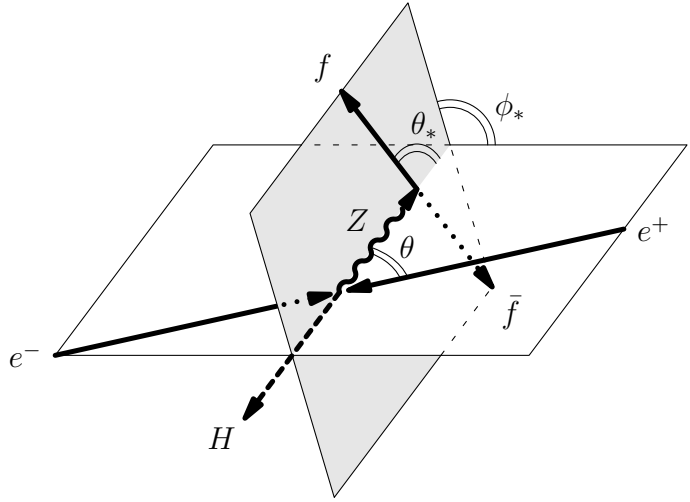}
\end{center}
\caption{Polar and azimuthal angles in the Higgs-strahlung
process. [The polar angle $\theta_*$ is defined in the $Z$ rest frame.]}
\label{angles}
\end{figure}

\paragraph{Total cross section and polar angular distribution.}
Denoting the polar angle between the $Z$ boson and the $e^+e^-$ beam
axis by $\theta$, the differential cross section for the process
$e^+e^-_{L,R}\to ZH$ may be written as
\begin{equation}
  \frac{d\sigma^{L,R}}{d\cos\theta}
        =  \frac{G_F^2 M_Z^4}{96\pi s}
           \left(v_e\pm a_e\right)^2\lambda^{1/2}\,
           \frac{\frac34\lambda\,\sin^2\theta\,\left(1+\alpha^{L,R}\right)
                 + 6\left(1+\beta^{L,R}\right)M_Z^2/s}
                {(1-M_Z^2/s)^2}
\end{equation}
and the integrated cross section
\begin{equation}
  \sigma = \frac{G_F^2 M_Z^4}{96\pi s}
           \left(v_e\pm a_e\right)^2\lambda^{1/2}
           \frac{\lambda\left(1+\alpha^{L,R}\right)
                 + 12\left(1+\beta^{L,R}\right)M_Z^2/s}
                {(1-M_Z^2/s)^2}
\end{equation}
The $Z$ charges of the electron are defined as usual by $a_e=-1$ and
$v_e=-1+4s_W^2$.  $s$ is the c.m.\ energy squared, and $\lambda$ the
two-particle phase space coefficient $\lambda =
\left[1-(m_H+m_Z)^2/s\right] \times\left[1-(m_H-m_Z)^2/s\right]$.  The 
coefficients $\alpha(s)^{L,R}$ and $\beta(s)^{L,R}$ can easily be 
determined for the interactions in Eqs.(\ref{L-ZZH}) and (\ref{L-contact}):
\begin{eqnarray}
  \alpha(s)^{L,R}
  &=& 2a_0 + (s-M_Z^2)\frac{8c_Ws_W}{v_e\pm a_e}c_{L,R}\\
  \beta(s)^{L,R}
  &=& \alpha(s)^{L,R} + 
      2\gamma\sqrt{s}\,M_Z
      \left[a_1 + \frac{4c_Ws_W}{v_e\pm a_e}
                  \left(1-\frac{M_Z^2}{s}\right)b_1\right]
\end{eqnarray}
where the boost of the $Z$ boson is given by $\gamma =
(s+M_Z^2-M_H^2)/2M_Z\sqrt{s}$.  

\begin{figure}[htbp]
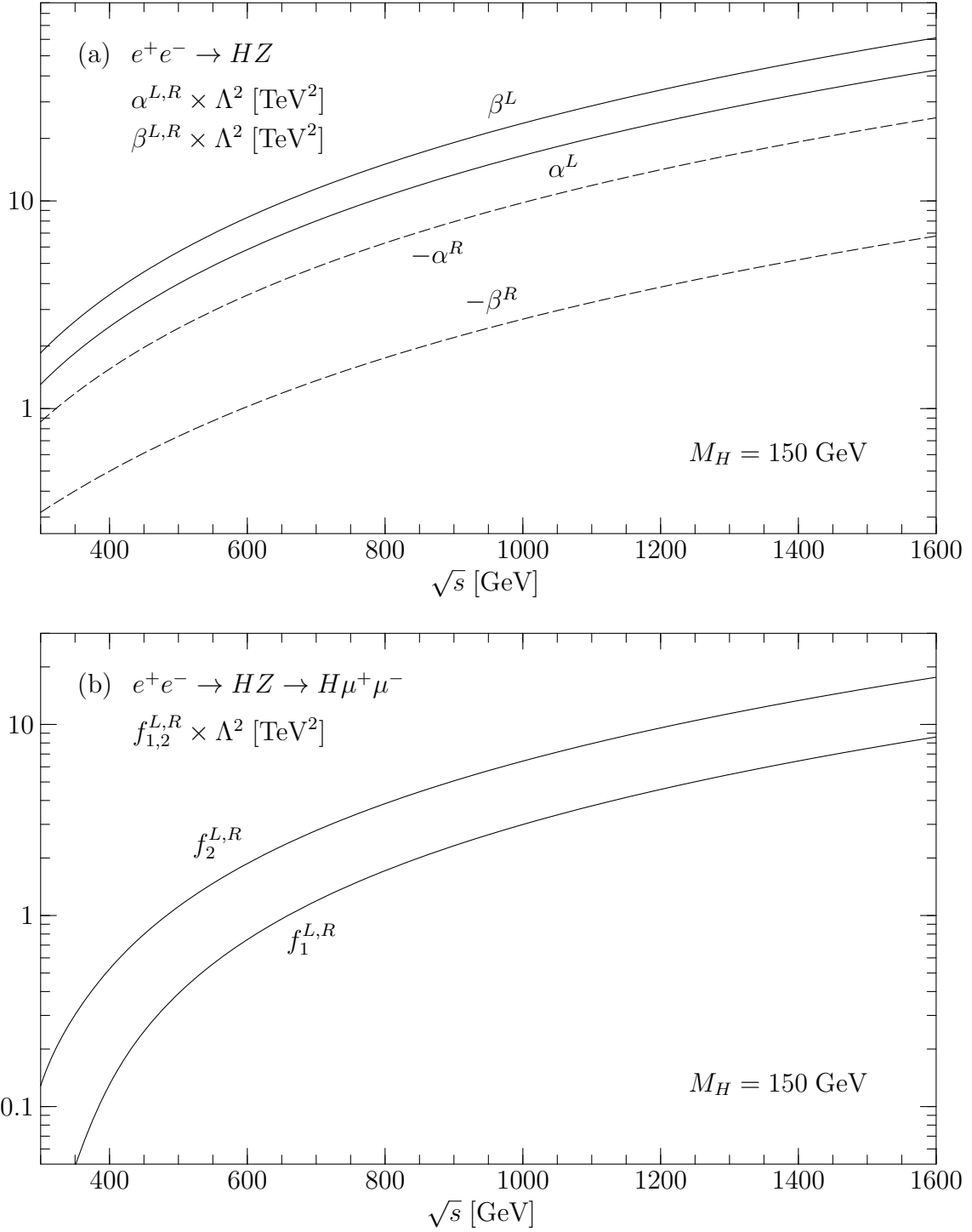

\begin{center}
\includegraphics{coeff.1}\\[1.5cm]
\includegraphics{coeff.2}\\[0.8cm]
\caption[dummy]{\label{coeff} Coefficients of the angular
  distributions as a function of the beam energy. Parameters are
  described in the text; in particular, $\alpha_i=1$ has ben chosen in
  the effective Lagrangian Eq.(\ref{L-eff}).  [The $L,R$ coefficients
  of the azimuthal distribution coincide for the special choice
  $\alpha_i = 1$.]}
\end{center}
\end{figure}

The modification of the cross section by the new interaction terms
has a simple structure.  The coefficient $a_0$ just renormalizes the
SM cross section.  By contrast, the contact interactions grow with
$s$.  [The ratio $s/\Lambda^2$ is assumed to be small enough for the
restriction to dimension-6 operators to be meaningful.]  The operators
$\op_{\varphi W}$, $\op_{\varphi B}$ affect the coefficient in the
cross section which is independent of $\theta$.  They damp the
fall-off of this term, changing the $1/s^2$ to a $1/s$ behavior;
however, these contributions remain subleading since they are
associated with transversely polarized $Z$ bosons which are suppressed
at high energies compared with the longitudinal components.  To
illustrate the size of the modifications $\alpha(s)^{L,R}$ and
$\beta(s)^{L,R}$, we have depicted these functions in
Fig.\ref{coeff}(a) for the special choice $\alpha_i = 1$.

\paragraph{Azimuthal distributions.}
The azimuthal angle $\phi_*$ of the fermion $f$ is defined as the
angle between the [$e^-,Z$\/] production
plane and the [$Z,f$\/] decay plane~(Fig.\ref{angles}).  It corresponds to the azimuthal angle of
$f$ in the $Z$ rest frame with respect to the [$e^-,Z$\/] plane.  On
general grounds, the $\phi_*$ distribution must be a linear function
of $\cos\phi_*$, $\cos 2\phi_*$, and $\sin\phi_*$, $\sin 2\phi_*$,
measuring the helicity components of the decaying spin-1 $Z$ state.
The coefficients of the sine terms vanish for CP invariant theories.
The $\cos\phi_*$ and $\cos 2\phi_*$ terms correspond to P-odd and
P-even combinations of the fermion currents.  The general azimuthal
distributions are quite involved~\cite{BZ,SH,GR}.  We therefore
restrict ourselves to the simplified case in which all polar angles
are integrated out, i.e., the polar angle $\theta$ of the $Z$ boson in
the laboratory frame and the polar angle $\theta_*$ of $f$ in the $Z$
rest frame. In this way we find for the azimuthal $\phi_*$ distribution:
\begin{equation}
  \frac{d\sigma^{L,R}}{d\phi_*} \sim 1 \mp\frac{9\pi^2}{32}\,
  \frac{2\,v_fa_f}{v_f^2+a_f^2}\,\frac{\gamma}{\gamma^2+2} 
  \left(1+f_{1}^{L,R}\right)\cos\phi_* + \frac{1}{2(\gamma^2+2)}
  \left(1+f_{2}^{L,R}\right)\cos 2\phi_*
\end{equation}
with 
\begin{eqnarray}
  f_1(s)^{L,R}
  &=& M_Z\sqrt{s}\,
         \frac{(\gamma^2-1)(\gamma^2-2)}{\gamma(\gamma^2+2)}
         \left[a_1 + \frac{4s_Wc_W}{v_e\pm a_e}
                     \left(1 - \frac{M_Z^2}{s}\right) b_1\right]
  \\
  f_2(s)^{L,R}
  &=& 2M_Z\sqrt{s}\,
         \frac{\gamma(\gamma^2-1)}{\gamma^2+2}
         \left[a_1 + \frac{4s_Wc_W}{v_e\pm a_e}
                     \left(1 - \frac{M_Z^2}{s}\right) b_1\right]
\end{eqnarray}
The cross section flattens with increasing c.m.\ energy in the
Standard Model, i.e.\ the coefficients of $\cos\phi_*$ and
$\cos2\phi_*$ decrease asymptotically proportional to $1/\sqrt{s}$ and
$1/s$, respectively.  The anomalous contributions modify this
behavior: The $\cos\phi_*$ term receives contributions which increase
proportional to $\sqrt{s}$ with respect to the total cross section,
while the $\cos2\phi_*$ term receive contributions from the
transversal couplings that approaches a constant value asymptotically.
The size of the new terms in $f_{1,2}^{L,R}$ is shown in
Fig.\ref{coeff}(b) as a function of the energy. [The special choice
$\alpha_i = 1$ we have adopted for illustration, implies $f_{1,2}^{L}
= f_{1,2}^{R}$.]

\paragraph{} It is instructive to study the high-energy behavior of
the coefficients in the limit $M_Z^2\ll s\ll\Lambda^2$.  In this case
we obtain the simplified relations:
\begin{eqnarray}
  \alpha(s)^{L,R} 
  &\simeq& \mp\, s\cdot 8s_Wc_W\, c_{L,R} + {\cal O}(v_e) \label{25}\\
  \beta(s)^{L,R} 
  &\simeq& \alpha(s)^{L,R}
           + s\left(a_1 \mp 4s_Wc_W\, b_1\right)+ {\cal O}(v_e) \label{26}
\end{eqnarray}
and
\begin{eqnarray}
  f_1(s)^{L,R}
  &\simeq& 
  \frac{s}{2}\left(a_1 \mp 4s_Wc_W\, b_1\right) + {\cal O}(v_e) 
  \label{27}\\
  f_2(s)^{L,R}
  &\simeq& s\left(a_1 \mp 4s_Wc_W\, b_1\right) + {\cal O}(v_e) \label{28}
\end{eqnarray}
Terms which are proportional to $v_e=-1+4s_W^2$ are suppressed by an
order of magnitude. If longitudinally polarized electrons are
available, the asymptotic value of the coefficients $a_1,b_1,c_L$ and
$c_R$ can be determined by measuring the polar angular distribution
without varying the beam energy.  The analysis of the azimuthal
$\phi_*$ distribution provides two additional independent measurements
of the coefficients $a_1$ and $b_1$. On the other hand, the set of
measurements remains incomplete for fixed energy if only unpolarized
electron/positron beams are used at high energies; in this case the
coefficients cannot be disentangled completely without varying the
beam energy within the preasymptotic region.

\baselineskip15pt
\begin{thebibliography}{19}
\bibitem{hst}   J.\ Ellis, M.K.\ Gaillard, and D.V.\ Nanopoulos,
                Nucl.\ Phys.\ {\bf B106} (1976) 292;
                B.L.\ Ioffe and V.A.\ Khoze,
                Sov.\ J.\ Part.\ Nucl.\ {\bf 9} (1978) 50;
                B.W.\ Lee, C.\ Quigg, and H.B.\ Thacker,
                Phys.\ Rev.\ {\bf D16} (1977) 1519;
                J.D.\ Bjorken, Proc.\ \emph{Summer Institute
                on Particle Physics}, SLAC Report 198 (1976).
\bibitem{LEP2}  M.\ Carena, P.M.\ Zerwas (conv.) et al.,
                \emph{Higgs Physics}, in: Proceedings of the
                Workshop
                \emph{Physics with LEP2},
                eds.\ G.\ Altarelli, T.\ Sj\"ostrand,
                and F.\ Zwirner, CERN Yellow Report 96-01.
\bibitem{FLC}   Proceedings of the Workshop
                \emph{$e^+e^-$ Collisions
                at 500 GeV: The Physics Potential},
                Munich--Annecy--Hamburg,
                ed.\ P.M.\ Zerwas,
                Reports DESY 92-123A,B; 93-123C.
\bibitem{BZ}    V.\ Barger, K.\ Cheung, A.\ Djouadi, B.\ Kniehl, 
                and P.M.\ Zerwas, 
                Phys.\ Rev.\ {\bf D49} (1994)~79.
\bibitem{BW85}  W.~Buchm\"{u}ller and D.~Wyler,
                Nucl.\ Phys.\ {\bf B268} (1986) 621.
\bibitem{GW}    B.\ Grzadkowski and J.\ Wudka,
                Phys.\ Lett.\ {\bf B364} (1995) 49.
\bibitem{SH}    K.\ Hagiwara and M.L.\ Stong,
                Z.\ Phys.\ {\bf C62} (1994) 99.
\bibitem{GR}    G.J.\ Gounaris, F.M.\ Renard, and N.D.\ Vlachos,
                Nucl.\ Phys.\ {\bf B459} (1996) 51;
                G.J.\ Gounaris, J.~Laissac, J.E.\ Paschalis,
                F.M.\ Renard, and N.D.\ Vlachos,
                Preprint PM/96-08.
\end{thebibliography}
\end{document}